\documentclass[11pt]{article}
\usepackage{citesort}
\usepackage{amsmath}
\usepackage{amssymb}
\usepackage[dvips]{graphicx}
\usepackage{psfrag}
\usepackage{xspace}
\newlength{\picwidth}
\newlength{\figurewidth}
\newcommand{\kbar}{\makebox[0pt][l]{\hspace{0.2ex}\rule[1.25ex]{0.8ex}{0.3pt}}k}
\newcommand{\unit}[1]{\mathit{#1}}
\newcommand{\half}{\tfrac{1}{2}}
\newcommand{\ks}{k}
\newcommand{\etal}{\textit{et al}\xspace }
\newcommand{\commutator}[2]{\left[#1,#2\right]}
\newcommand{\abs}[1]{\left|#1\right|}
\newcommand{\subtext}[1]{\mbox{\scriptsize #1}}
\newcommand{\PDM}{\hat{\mathcal{P}}}  
\newcommand{\ket}[1]{\left|#1\right\rangle}
\newcommand{\bra}[1]{\left\langle#1\right|}

\newcommand{\Hdark}{H_{\text{dark}}}
\newcommand{\Hlight}{H_{\text{light}}}
\begin{document}
\title{A study of quantum decoherence in a system with Kolmogorov-Arnol'd-Moser tori}

\author{G. H. Ball, K. M. D. Vant, H. Ammann and N. L. Christensen\\ Department of Physics, University of Auckland, Auckland, New Zealand}
\maketitle

\begin{abstract}
We present an experimental and numerical study of the effects of
decoherence on a quantum system whose classical analogue has
Kolmogorov-Arnol'd-Moser (KAM) tori in its phase space. Atoms are
prepared in a caesium magneto-optical trap at temperatures and
densities which necessitate a quantum description. This real
quantum system is coupled to the environment via spontaneous
emission. The degree of coupling is varied and the effects of this
coupling on the quantum coherence of the system are studied. When
the classical diffusion through a partially broken torus is
$\lesssim\hbar$, diffusion of quantum particles is inhibited. We
find that increasing decoherence via spontaneous emission
increases the transport of quantum particles through the boundary.
\end{abstract}
\section{Introduction}
\label{intro}
The study of decoherence in a quantum system has been a subject of
much interest in recent years.  Since the emergence of quantum
mechanics some seventy years ago, a central problem in its
interpretation has been the fact that the linearity of the
Schr\"{o}dinger equation allows macroscopic physical states to be
derived from arbitrary superpositions of other states, a situation
which is quite clearly in contradiction to our experience.  This
was originally illustrated in the familiar paradox of
Schr\"{o}dinger's cat. One simple explanation is to suppose that
an initially classical state for a macroscopic system will always
evolve into another classical state so that for a universe with a
suitable initial condition, non-classical states occur only when
carefully prepared by some experimentalist. In this interpretation
Schr\"{o}dinger's cat presents no paradox, but merely lies outside
our normal realm of experience. This hypothesis can be rejected in
the light of quantum analysis of real macroscopic, classically
chaotic systems. Calculations (see for example Zurek's discussion
in terms of celestial objects \cite{Zurek3}) indicate that in
these systems, quantum dynamics differ from classical after
relatively short times, and lead to flagrantly non-classical
states.  Any successful theory must explain why these states are
not found in practice.

There have been several more sophisticated explanations as to why
quantum states are not evident in our everyday world. The first
was Neils Bohr's Copenhagen Interpretation (CI) \cite{Bohr}. Bohr
theorized that quantum theory was not universal --- a classical
apparatus was necessary to carry out the measurements. He also
required that the boundary between quantum and classical domains
was mobile and could be pushed back by appropriate apparatus.
Another significant school of thought was Hugh Everett's Many
World's Interpretation (MWI) \cite{Everett,Wheeler}. He proposed
that each time an interaction which would produce a superposition
takes place, the wave-function of the universe bifurcates to give
an ever increasing number of `branches'. In 1952, David Bohm
\cite{Bohm} detailed an interpretation based on `hidden' variables
which had previously been investigated to some extent by
de~Broglie \cite{deBroglie}. Bohm theorized that these hidden
variables would allow ``a detailed causal and continuous
description of all processes, and not require us to forego the
possibility of conceiving the quantum level in precise terms.''
Many other approaches to the problem have been taken, examples of
which can be found in \cite{Wheeler2,Power,Penrose}.

A more recent and arguably more satisfactory solution to the
problem of quantum-classical correspondence has been championed by
Wojciech Zurek and co-workers \cite{Zurek1,Zurek2}. They suggest
that any real, open quantum system leaks coherence to its
surroundings via extraneous degrees of freedom which are coupled
to the environment.  A suitable environment might be provided by
air molecules, cosmic background microwave photons, or the vacuum
point fluctuations of the electromagnetic field. The rate at which
this decoupling proceeds depends on the particular state of the
system, the dynamics of the system and the form of the interaction
with the environment.  The states most resistant to decoherence
form a (not necessarily orthogonal) basis of `pointer' states
which are `selected'; the density matrix rapidly becomes diagonal
in this basis \cite{Zurek4}. Any particular realization of the
system quickly evolves into one of these basis states, and only
measurements in this basis can be thought of as revealing the true
state of the system.  It appears to be a general result that the
``pointer'' states are the ``most classical'' states of the
system. In most cases for macroscopic systems the pointer states
are expected to be those which are localized to a high degree.
This arises from the fact that in most cases the system's
interactions with the environment are chiefly position dependent.
Superpositions of the pointer states of a system will collapse on
a time scale determined by the coupling. In a measurement of a
microscopic system by a macroscopic observer, the wave-function
collapse of traditional measurement theory is caused by
decoherence of the meter entangled with the system.  Bizarre
macroscopic quantum states might in theory be prepared, but will
survive only for a vanishingly short time, and the classical
description of the world as we observe it will be regained.
Quantum mechanics without fundamental modification retains its
position as the true description of the universe, and the limited
set of states which we observe around us is accounted for.

In this paper, we study experimentally the effects of increasing
the coupling of a quantum system to its environment. Our
experimental system is a low density cloud of super cool ($\sim 15
\unit{\mu K}$) caesium atoms prepared in the ultra high vacuum
glass cell of a magneto-optical trap (MOT). Because of the low
densities achieved in a MOT, the interactions of the atoms are
negligible and each atom can be modeled as a quantum particle in a
periodic potential (see Section \ref{ghb1}). With $10^{5}$ atoms
trapped per experimental run, we deal with statistically
significant numbers of particles so the experimental distributions
will closely approximate quantum probability distributions. We
temporally modulate a standing wave optical potential which then
creates Kolmogorov-Arnol'd-Moser (KAM) tori (impenetrable momentum
barriers) in the classical phase space of the system. For an
increased Rabi frequency (i.e.\ increased perturbation), holes
appear in the barriers, now called cantori, through which the
atoms can diffuse. The diffusion rates of classical and quantum
particles are distinctly different, with quantum diffusion being
largely suppressed by the cantori. The decohering effects of
coupling to the environment bring the behaviour of the quantum
ensemble towards the classical limit.
\section{The Kolmogorov-Arnol'd-Moser Theorem}
For a closed system with an integrable Hamiltonian, all solutions
lie on two dimensional tori embedded in $2N-1$ dimensional phase
space, where $N$ is the number of degrees of freedom of the
system. Each solution or trajectory is indefinitely confined to
its own torus. The motion of a trajectory on a torus can be
described by two coordinates, $\theta_{1}$ and $\theta_{2}$ on
$[0, 2\pi)$. The winding number $w$ of the torus is defined as
\begin{equation}
w = \frac{ \omega_{1} }{ \omega_{2} },\;
\mathrm{where}\;\omega_{j} = \dot{\theta}_{j},\;\mathrm{for}\;
j=1,2.
\end{equation}
If the two frequencies $\omega_{j}$ are commensurate then the
winding number of the torus is rational, and trajectories on the
torus are periodic. If the frequencies are incommensurate the
winding number is irrational, and the motion in phase space is
quasi-periodic.

When the system Hamiltonian is perturbed such that it becomes
non-integrable, non-linear resonances appear in the phase space at
the location of tori with rational winding numbers, altering the
topology and destroying these tori (see Section~\ref{ghb1}).
Trajectories which were confined on the vanished tori, now
traverse the same general region of phase space.  However the KAM
theorem states that tori with irrational winding numbers (KAM
tori) are not immediately destroyed by small amounts of
non-integrability and continue to act as tori in the new phase
space.  For increasing perturbation, the non-linear resonances
grow and the KAM tori are eventually destroyed by nearby
resonances.  Within a given region of phase space, those with the
most irrational winding numbers are the last to break up.
Irrationality is measured by the rate of convergence of the
continued fraction representation,
\begin{equation}
w\equiv[a_{0},a_{1},a_{2},\ldots]=a_{0}+\cfrac{1}{a_{1}+
                                         \cfrac{1}{a_{2}+
                                          \cfrac{1}{a_{3}+\ldots
                                          }}}
\end{equation}

When a KAM torus is destroyed, a \emph{cantorus} is left in its
place. The properties of a broken torus or cantorus in a quantum
system have been the subject of considerable interest and several
numerical studies \cite{Geisel,Geisel2,Brown,MacKay,MacKay2}.
\section{Our analytical system}
\label{ghb1}
\begin{figure}[p]
\psfrag{aT}{\large{$T\alpha$}}\psfrag{dT}{\large$T\Delta$}
\psfrag{T}{\large$T$} \centering
\includegraphics[clip,width=\textwidth,bb=78 107 456 287]{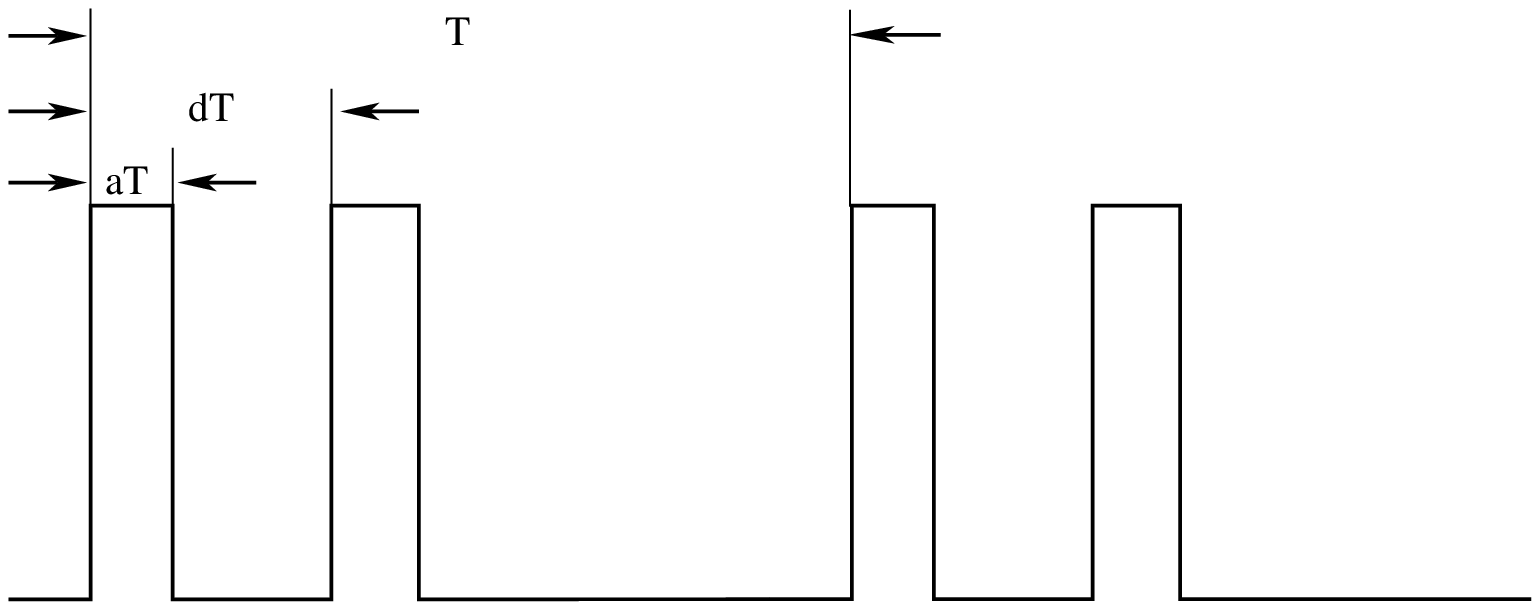}
\caption{Double pulse showing definitions of $\alpha$ and
$\Delta$}
\label{doublepulsedef}
\end{figure}
The atom interacts with a standing wave of near-resonant light
(frequency $\omega_{L}$) which  is temporally modulated with
period $T$.  When the detuning $\delta_{L}=\omega_{0}-\omega_{L}$
(where $\omega_{0}$ is the resonant frequency of the transition)
is sufficiently large compared to the resonant Rabi frequency
$\Omega/2$ (proportional to the square root of the standing wave intensity),
the amplitudes of the excited states can be adiabatically
eliminated. The dynamics are governed by stimulated two-photon
scattering between ground states, with momentum changes in units
of $2\hbar k_{L}$. Classically the atom behaves as a dipole in a
conservative one dimensional potential.  The Hamiltonian in this
limit is given by
\begin{equation}
\mathcal{H}=\frac{p_{x}^{2}}{2M}- \frac{\hbar
  \Omega_{\subtext{eff}}}{8} \cos 2k_{L}x
\sum_{n=-\infty}^{\infty}f\left(\frac{t}{T}-n\right)
\end{equation}
where $f(t/T)$ specifies the temporal shape of the ``kicks,''
(with $0\leq f(t/T)\leq 1$), $k_{L}$ is the laser wave number, and
$\Omega_{\subtext{eff}}$ is the effective Rabi frequency.  For a
two-level atom $\Omega_{\subtext{eff}}=\Omega^{2}/\delta_{L}$ but
for our system we instead have $\Omega_{\subtext{eff}}=
\Omega^{2}(s_{45}/\delta_{45}+s_{44}/\delta_{44}+s_{43}/\delta_{43})$,
where the terms in brackets take into account the different dipole
transitions between the relevant hyperfine levels in caesium
($F=4\rightarrow F'=5,4,3$).  In our simulations we assumed equal
populations of the Zeeman sub-levels, yielding numerical values for
the $s_{4j}$ of $s_{45}=11/27$, $s_{44}=7/36$, and $s_{43}=7/108$;
$\delta_{4j}$ are the corresponding detunings.  Note that the
different magnetic sub-levels will experience different AC Stark
shifts.  For the smallest detuning used in this work this results
in a 5\% spread in the coupling strength.

We can write the Hamiltonian in dimensionless form as
\begin{equation}\label{eq:dimensionless hamiltonian}
  H=\frac{\rho^{2}}{2}-\ks \cos \phi \sum_{n=-\infty}^{\infty}f(\tau-n)
\end{equation}
where $\ks =\hbar\Omega_{\subtext{eff}}k_{L}^{2}T^{2}/2M$ is a
measure of the perturbation of the system called the kick
strength, $\tau=t/T$, $\phi=2k_{L}x$, $\rho=(2k_{L}T/M)p_{x}$, and
$H=(4k_{L}^{2}T^{2}/M)\mathcal{H}$. If we consider the dynamics on
a single half wavelength of the laser beam, the Hamiltonian is
that of a driven rotor, with dimensionless parameters $I=1$ and
$\omega_{0}^{2}=\ks$, where $\omega_{0}$ is the small amplitude
oscillation frequency of the rotor.  In the quantized model,
$\phi$ and $\rho$ are conjugate variables with a commutation
relation $\commutator{\rho}{\phi}=-i\kbar$, where $\kbar=4\hbar
k_{L}^{2}T/M$ is our scaled Planck constant.

Experimentally, we use a double pulse kick (see
Figure~\ref{doublepulsedef} and Section~\ref{exptsection}). For
this $f(\tau)$ the Hamiltonian can alternatively be written as
\begin{equation}
H=\frac{\rho^{2}}{2}-\ks\sum_{m=-\infty}^{\infty}a_{m}\cos(\phi-2\pi m\tau)
\end{equation}
where
$a_{m}=\frac{1}{10}\mathrm{sinc}\,\frac{m\pi}{20}\cos\frac{m\pi}{10}$.
(With the sinc function defined as sinc$(x)=\sin(x)/x$).  The
rotor is driven by traveling cosine waves, the speed of the
$m^{th}$ cosine wave corresponding to a dimensionless momentum of
$\rho=2\pi m$. There will be a primary resonance in the phase
space wherever the speed of rotation of the rotor matches the
speed of a cosine wave \cite{reichl}. In the reference frame in
which the cosine wave is at rest, the rotor will be trapped in a
pendulum potential. This causes a change in the topology of the
classical phase space at $\rho=\rho_{m}$, for $a_{m}\ne 0$. For
$m$ such that $a_{m}=0$ there is a missing primary resonance. The
widths of the primary resonance zones are given by
$\delta\rho_{m}=4\sqrt{a_{m}\ks}$. The Chirikov condition for
overlap is that the spacing between the resonances be equal to the
sum of the half-widths or
$\abs{\rho_{m}-\rho_{n}}=2\sqrt{a_{m}\ks}+2\sqrt{a_{n}\ks}$.
Overlap of adjacent resonances is the mechanism for the
destruction of KAM tori. We expect tori in the vicinity of missing
resonances to survive higher kick strengths than all others. A
more sophisticated analysis must take into account the appearance
of secondary resonances arising from interactions between the
primary resonances.
\section{Experimental Set-up}
\label{exptsection}
Our experimental setup is very similar to that used in the
experiments of Ammann \etal \cite{Ammann1,Ammann2} and Moore \etal
\cite{Raizen1}. Approximately $10^{5}$ caesium atoms are trapped
and laser cooled in a standard MOT powered by two diode lasers
operating in the infra-red ($852\unit{nm}$).  The initial trapped
cloud has a FWHM of $\sim 200\unit{\mu m}$ and a temperature of
$10-15\unit{\mu K}$. The periodically modulated potential
(described in Section \ref{ghb1}) is provided by a third diode
laser. The beam from this laser passes through an $80\unit{MHz}$
acousto-optic modulator (AOM) and a single mode optical fibre
which spatially filters the light and delivers it to the trap. The
beam is then collimated and retroreflected from a mirror on the
opposite side of the trap to form a standing wave potential across
the atomic cloud.  The calculated beam waist at the cloud is
$795\unit{\mu m}$. This potential is temporally modulated via the
rf supply to the AOM. Our calculated Rabi frequency at the centre
of the trap, for maximum optical power is $\Omega/2\pi=310\
\unit{MHz}$.  A reasonably narrow distribution in the kicking
strength $\ks$ is produced by the finite widths of the cloud and
the beam waist of the kicking potential (RMS spread of 6\% and
$\ks_{mean}\approx 0.94\ks_{max}$). In the remainder of this
paper, $\ks$ always refers to $\ks_{mean}$.

\begin{figure}
\centering
\includegraphics[bb=140 120 360 600]{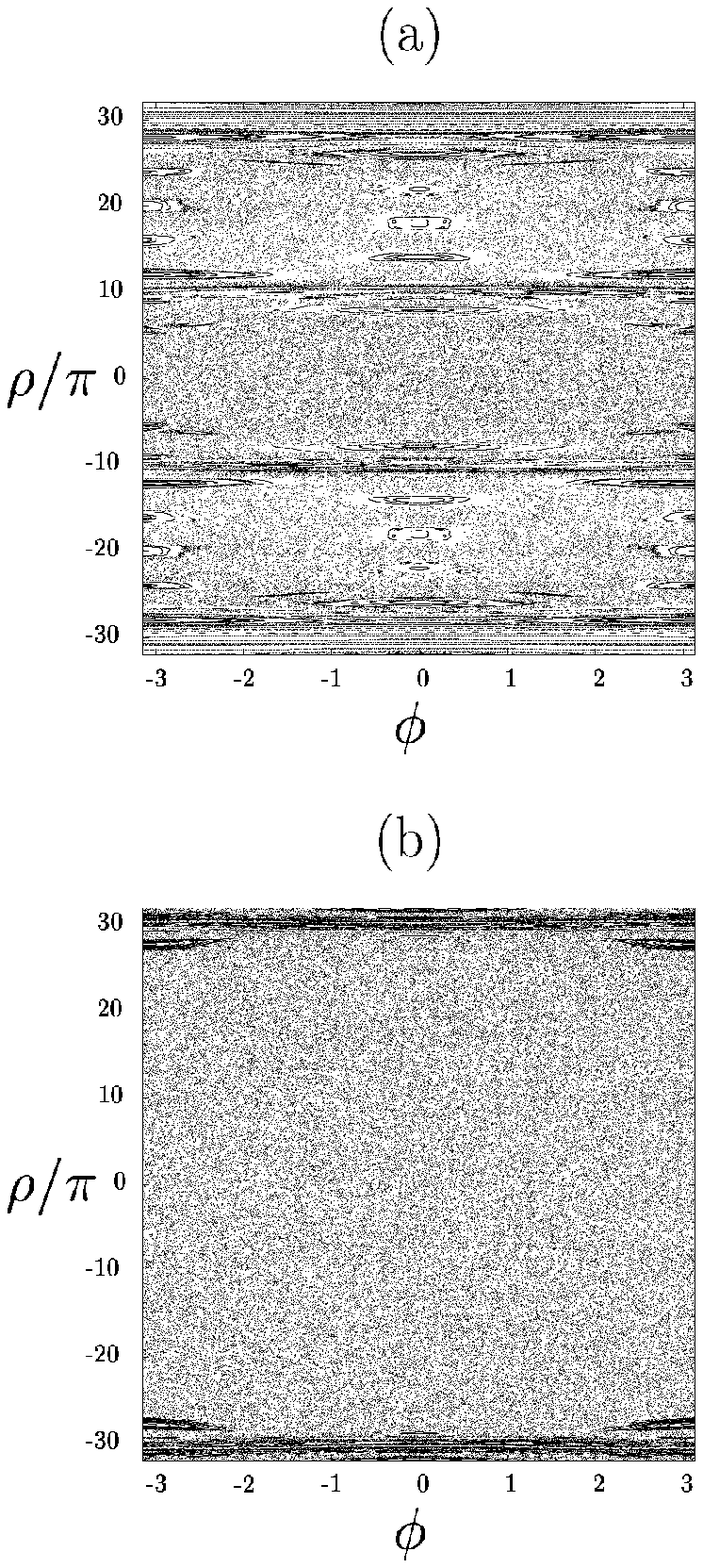}
\caption{(a) is the Poincare section of the system for a very low
kick strength.  The unbroken KAM tori are clearly visible at $\rho
= \pm10\pi\;\mathrm{and}\;\pm30\pi$,  (b) shows the same Poincare
section for $\ks \sim 300$ where the cantori at $\rho=\pm10\pi$
have completely vanished from the phase space} \label{pc}
\end{figure}

In the system Hamiltonian, the pulse train is described by
$f(\tau)$. Any pulse of finite length will create KAM tori in the
phase space of the system. These impenetrable barriers to
diffusion were observed in recent quasi $\delta$-kicked rotor
experiments \cite{Ammann1,Ammann2,Raizen3} and were discussed in a
recent paper by Klappauf \etal \cite{Klappauf}. In these studies
the effects of the boundaries were avoided by tailoring the pulse
length to push them into regions of phase space beyond the
localization length of dynamical localization. However, they
present a very interesting area of study in themselves and have
been investigated both numerically and experimentally
\cite{Geisel,Geisel2,Brown,MacKay,MacKay2,Christensen,Vant}. A
pulse train consisting of single pulses is not the best system for
studying KAM tori because the only energetic chaotic sea lies
within the tori of lowest momentum. Energetic chaotic seas aid the
diffusion of particles away from the partially permeable barrier
and are important experimentally in isolating the effect of the
boundary in phase space. We use a double pulse per kicking cycle
where the pulse period is $T = 25 \unit{\mu s}$ with pulse width
$\alpha = 1/20$ and pulse spacing $\Delta = 1/10$ (see
Figure~\ref{doublepulsedef}). This pulse shape gives energetic
chaotic seas on both sides of the long-lived KAM tori with the
smallest absolute momentum (see Figure~\ref{pc}).
To achieve varying levels of spontaneous emission, we varied
$\delta_{L} = \omega_{0}-\omega_{L}$, the detuning of the kicking
potential from resonance, while simultaneously altering the beam
intensity to maintain a constant kicking strength.
\section{Transport through cantori}
Classical particles which are no longer confined to their own tori
can diffuse rapidly even through a recently broken cantorus. The
cantorus no longer forms an impenetrable barrier in the classical
phase space and trajectories from both sides of the original
boundary can eventually fill the entire region between two
adjacent, unbroken tori. For quantum particles however, a very
different behaviour had been predicted. Various numerical studies
of particle transport through cantori
\cite{Geisel,Geisel2,Brown,MacKay,MacKay2} suggested that until
the phase space area escaping through the cantori per kick cycle
was $\sim\hbar$, the diffusion of quantum particles would be
restricted to a quantum tunnelling like behaviour. Hence the
prediction was for a distinct difference between the diffusion
rates of classical and quantum particles. Figure~\ref{pc}a shows a
Poincare section of our system for a very low kicking strength.
The KAM tori at $\rho = \pm10\pi$ and $\pm30\pi$ are clearly
visible as regions of stability between large chaotic seas.
Figure~\ref{pc}b shows the Poincare section for $\ks \sim 300$,
which is comparable to the kicking strengths used in our
experimental work. The tori at $\rho=\pm10\pi$ are no longer
visible in the phase space at this resolution; however they have
been shown to still have a significant effect on the dynamics of
both the quantum \cite{Christensen,Vant} and classical systems.

\begin{figure}[p]
\psfrag{l}[rt][cc]{\large 0}   \psfrag{m}[rt][cc]{\large 50}
\psfrag{n}[rt][cc]{\large -20} \psfrag{o}[rt][cc]{\large 0}
\psfrag{p}[rt][cc]{\large 20}

\psfrag{(a)}[lt][lB]{\huge (a)} \psfrag{(b)}[lt][lB]{\huge (b)}
\psfrag{(c)}[lt][lB]{\huge (c)} \psfrag{(d)}[lt][lB]{\huge (d)}
\psfrag{x}[lt][lB]{\huge$\rho/\pi$}
\psfrag{y}[rt][lB]{\huge$\tau$} \centering
\includegraphics[width=\textwidth]{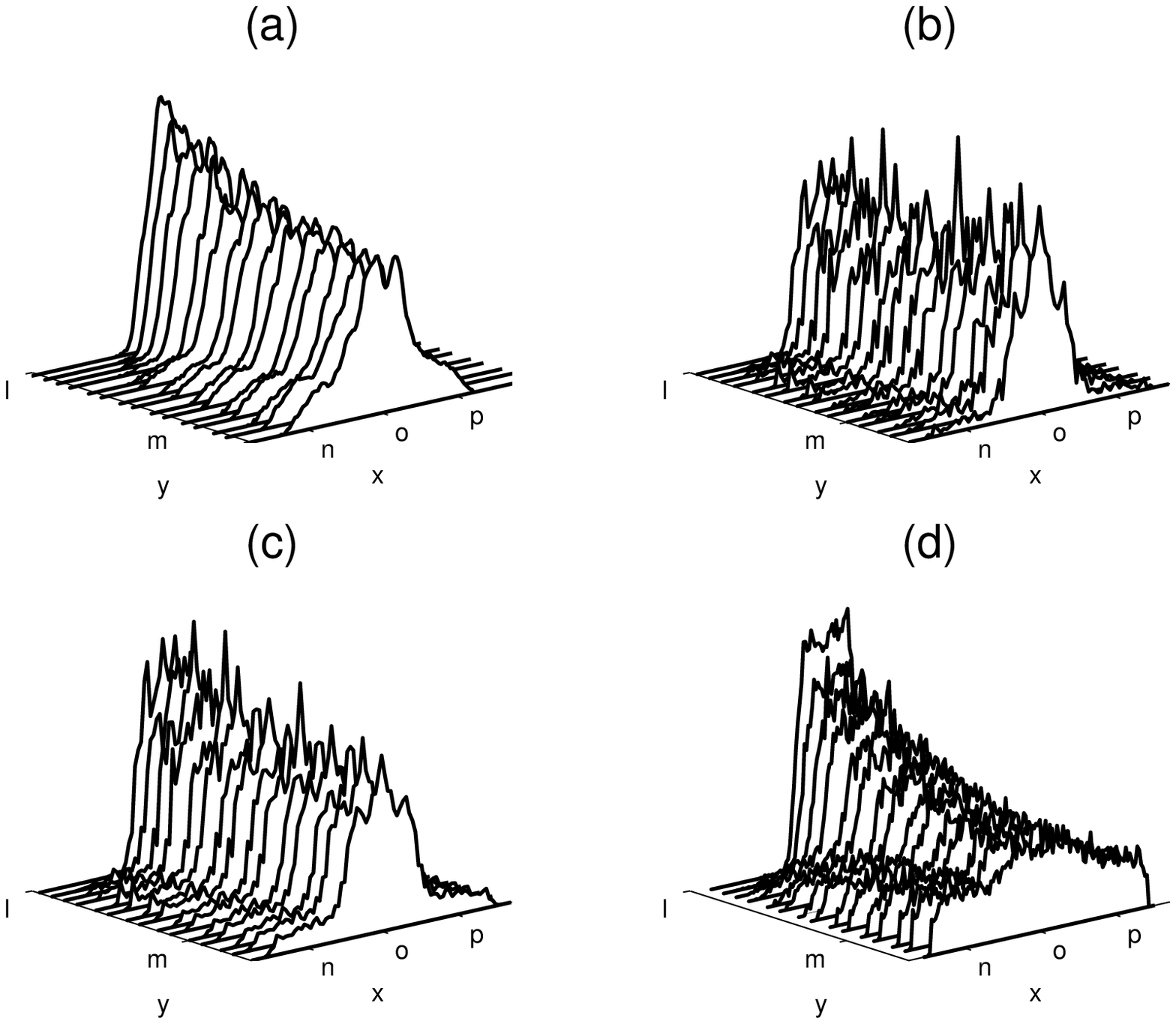}
\vspace{30pt} \caption{Momentum distributions for a kick strength
$\ks=280$ as a function of the number of kicking cycles $\tau$.
The experimental data (a) with its probability of spontaneous
emission per cycle $\eta=1.9\%$, shows its distinctive shoulders
at $\rho=\pm10\pi$. (b) is a quantum simulation for the
experimental parameters with no spontaneous emission, (c) is a
quantum simulation including spontaneous emission $\eta=1.9\%$ and
(d) shows the corresponding classical simulation.}
\label{waterf}
\end{figure}

We prepare our atoms so that they initially lie within the
$\rho=\pm 10\pi$ cantori and monitor their subsequent evolution
through the boundary. The final momentum distributions of the
atomic cloud display distinctive `shoulders' at $\rho=\pm10\pi$
(as shown in Figure~\ref{waterf}) due to the inhibition of
diffusion introduced by the cantori.
This KAM localization is distinct from the more widely studied
`dynamical localization'.  The signature of dynamical localization
is an exponential lineshape in momentum space which is markedly
different from the box-like distributions and characteristic
shoulders observed with KAM boundaries.  Also, for this
experiment, the localization length of the system, $l_{\rho} \sim
170$ is considerably longer than the momentum width of the
$1^{st}$ and $2^{nd}$ KAM boundaries at $\rho=\pm10\pi$ and
$\pm30\pi$ and hence KAM localization occurs before dynamical
localization can have an effect.
\section{Decoherence}
The theory of decoherence provides the most recent and so far
perhaps the most satisfying of a series of attempts to explain the
disparities between the predictions of quantum mechanics and the
everyday experiences of the world we inhabit (see
Section~\ref{intro}). The field of quantum chaos provides an ideal
backdrop for a study of the predicted effects of decoherence. It
is now widely accepted that sensitive dependence on initial
conditions - the hallmark of classical chaos - does not occur in
closed quantum systems.  This raises problems for the
quantum-classical correspondence (QCC) principle. Quantum
mechanics must be able to describe the classical limit of chaotic
behaviour. In this case, employing the limit as $\hbar \rightarrow
0$ is not entirely satisfactory. Chaotic systems can develop
highly complex phase space structures in logarithmically short
times and hence the small but non-zero value of $\hbar$ is an
important factor.

According to the work of Zurek \etal \cite{Zurek1,Zurek2}, these
difficulties in restoring the classical behaviour can be
eliminated by realizing that it is not possible to isolate
macroscopic quantum systems from their environment.  The coupling
of the extraneous degrees of freedom of a system to the
environment destroys the quantum coherences on a time scale
inversely proportional to the degree of coupling. In our
experiments and simulations, we introduce coupling via spontaneous
emission induced by the kicking potential. As the level of
spontaneous emission increases so does the coupling of the system
to the vacuum fluctuations which constitute the environment. The
predicted effect of this increasing coupling is an increase in the
transport of quantum particles across the cantori as the quantum
diffusion rate tends to its classical limit.
\begin{figure}[p]
\centering
\includegraphics[bb= 150 200 430 650]{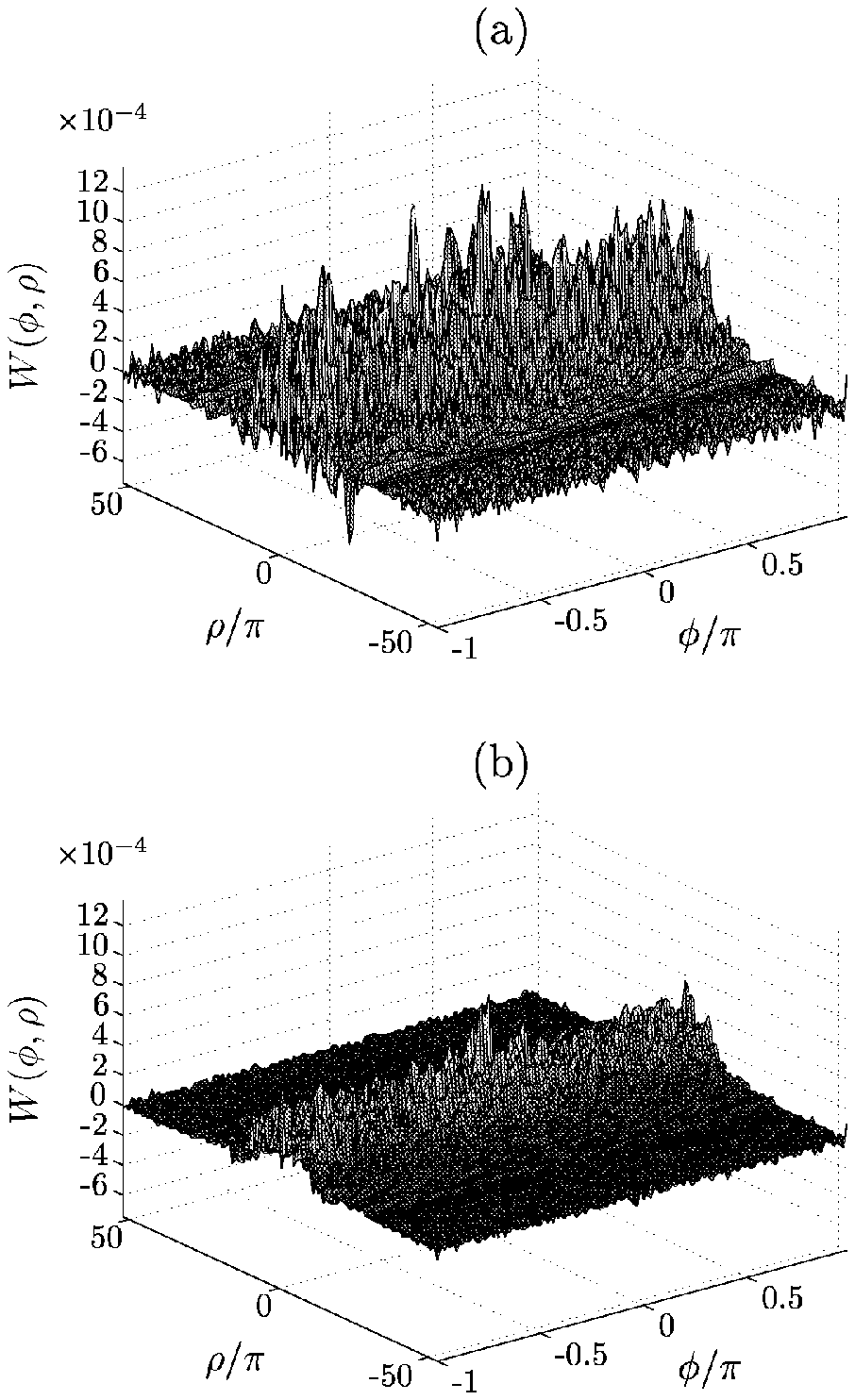}
\caption{Wigner functions for the system after 70 kicks, with
$\ks=280$. (a) corresponds to a probability of spontaneous
emission per kick cycle $\eta=0\%$, giving pure quantum evolution.
We see rapid fluctuations and negative regions, indicating quantum
interference effects.  (b) shows results for $\eta=2\%$,
comparable to experimental value. Quantum effects are reduced
leading to a more classical distribution.} \label{3dwigner}
\end{figure}
\label{ghb2}
For the purposes of simulation
we note that $H=\rho^{2}/2-\ks\cos\phi=\Hlight$ while the driving
potential is `switched on', and $H=\rho^{2}/2=\Hdark$ otherwise.
Classical trajectories involve periods of pendulum motion
described by Jacobi elliptic functions, alternating with periods
of free evolution. Usually $10^{4}$ trajectories are followed
during each run.  For each pulse, the amplitude and phase of the
elliptic function is matched to the position and motion of each
trajectory by inverting the elliptic function numerically.  To
simulate our experiments, we choose a thermal random distribution
of initial conditions.  To produce Poincare sections, more uniform
conditions are chosen, and optimized to reveal the phase space
structure.

The quantum system is represented by a basis of $N=128$ momentum
eigenstates $\ket{n}$, where $\rho\ket{n}=n\kbar\ket{n}$ and
$n=-64,\ldots,63$. The evolution operator for a single kick is
\begin{multline}
U=\exp(-i\frac{17\Hdark}{40\kbar})
  \exp(-i\frac{\Hlight}{20\kbar})
  \exp(-i\frac{\Hdark}{20\kbar})\ldots\\
\times  \exp(-i\frac{\Hlight}{20\kbar})
  \exp(-i\frac{17\Hdark}{40\kbar})
\end{multline}
where $\bra{m}\Hdark\ket{n}=\half n^{2}\kbar^{2}\delta_{m,n}$,
$\bra{m}\Hlight\ket{n}=\half n^{2}\kbar^{2}\delta_{m,n}-\half\ks
(\delta_{m,n+1}+\delta_{m,n-1})$ and we treat $n$ as periodic. The
$N\times N$ matrices are exponentiated numerically.  In the case
of evolution without spontaneous emission, the evolution is
entirely coherent, and is solved by finding the eigenvectors of
the evolution operator (Floquet method). For incoherent evolution,
the effects of spontaneous emission are simulated by adding the
density matrix to two shifted versions of itself, once per kick:
\begin{equation}
\bra{m}\PDM\ket{n}=\half\eta\left(\bra{m+1}\PDM\ket{n+1}+
  \bra{m-1}\PDM\ket{n-1}\right)+(1-\eta)\bra{m}\PDM\ket{n}
\end{equation}
where $\PDM$ is the density operator and $\eta$ is the probability for
a particular atom to spontaneously emit during one kick cycle.

A convenient way to visualize the information represented by the
density matrix is in the form of a Wigner function.  For a
discrete, truncated basis we use the toriodal Wigner function as
defined in \cite{kolovsky:96}.
\begin{equation}
w(X_{k},P_{l},t)=\sum_{j=0}^{2N-1}\exp\left(i\frac{\pi jk}{N}\right)
\frac{1+(-1)^{l+j}}{2}\bra{\frac{l+j}{2}}\PDM\ket{\frac{l-j}{2}}
\end{equation}
Where $P_{l}=(\kbar/2)l$ and $X_{k}=\pi k/N$.  This gives a Wigner
function defined on a $2N\times 2N$ grid.  Averaging over cells of
four adjacent points we reduce the grid to $N \times N$. Examples
of results obtained are shown in Figure~\ref{3dwigner}. We see
that decoherence smooths the Wigner function, removing the rapid
oscillations and negative regions which are characteristic of
quantum interference phenomena.
\section{Experimental results}
\begin{figure}[p]
\psfrag{a}[c][c]{\large 0} \psfrag{b}[c][c]{\large 10}
\psfrag{c}[c][c]{\large 20} \psfrag{d}[c][c]{\large 30}
\psfrag{e}[c][c]{\large 40} \psfrag{f}[c][c]{\large 50}
\psfrag{g}[c][c]{\large 60} \psfrag{h}[c][c]{\large 70}
\psfrag{i}[rB][rB]{\large 0} \psfrag{j}[rB][rB]{\large 10}
\psfrag{k}[rB][rB]{\large 20} \psfrag{l}[rB][rB]{\large 30}
\psfrag{m}[rB][rB]{\large 40} \psfrag{n}[rB][rB]{\large 50}
\psfrag{o}[rB][rB]{\large 60} \psfrag{p}[rB][rB]{\large 70}
\psfrag{x}[ct][cB]{\Large Number of double kicks}
\psfrag{y}[cB][ct]{\Large \% outside $1^{st}$KAM boundary}
\centering
\includegraphics[width=\textwidth]{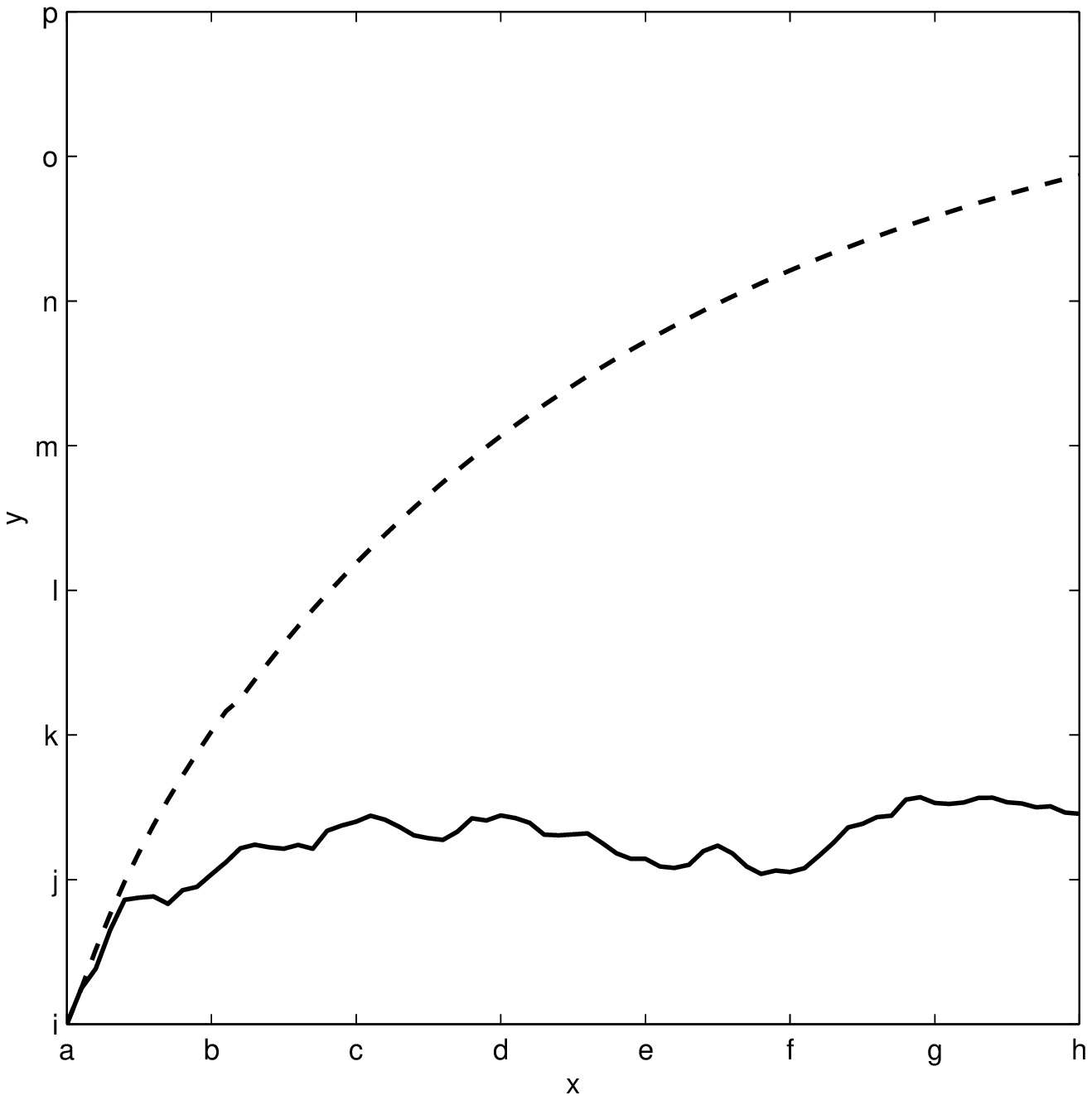}
\vspace{30pt}
\caption{Quantum (solid) and Classical (dashed)
simulations for a kicking strength of $\ks=270$ showing the
percentage of atoms to cross the $\rho=\pm10\pi$ cantori. This
graph clearly shows the inhibition to quantum diffusion presented
by the cantorus.}
\label{quantclas}
\end{figure}

In the absence of decoherence the caesium atoms behave as quantum
particles. Thus even for cantori where significant diffusion of
classical particles can occur, our quantum particles should still
be strongly contained as can be seen from our classical and
quantum simulations in Figure~\ref{quantclas} where we calculate
the percentage of atoms that cross the $\rho=\pm10\pi$ cantori as
a function of the number of kick cycles.
Not until the phase space escaping through the cantorus per cycle
is $\sim\hbar$, or in our scaled units $\sim\kbar$, do the quantum
particles move to any great extent across the boundary. For our
experimental parameters, the phase space flux through the
$\rho=\pm10\pi$ cantori per kicking cycle is $\sim 4.6\kbar$ so
the quantum diffusion will be strongly inhibited. The predicted
effect of increasing the coupling to the environment then is to
increase the transfer of atoms across the cantorus. As the quantum
system becomes more and more strongly coupled to its environment,
the behaviour of the atoms is expected to approach the classical
limit of rapid diffusion.

\begin{figure}[p]
\psfrag{a}[c][c]{\large 0} \psfrag{b}[c][c]{\large 10}
\psfrag{c}[c][c]{\large 20} \psfrag{d}[c][c]{\large 30}
\psfrag{e}[c][c]{\large 40} \psfrag{f}[c][c]{\large 50}
\psfrag{g}[c][c]{\large 60} \psfrag{h}[c][c]{\large 70}
\psfrag{k}[rB][rB]{\large 0} \psfrag{n}[rB][rB]{\large 10}
\psfrag{o}[rB][rB]{\large 20} \psfrag{p}[rB][rB]{\large 30}
\psfrag{q}[rB][rB]{\large 40} \psfrag{r}[rB][rB]{\large 50}
\psfrag{s}[rB][rB]{\large 60} \psfrag{t}[rB][rB]{\large 70}

\psfrag{x}[ct][cB]{\Large Number of double kicks}
\psfrag{y}[cB][ct]{\Large \% outside $1^{st}$KAM boundary}
\centering
\includegraphics[width = 1.1\textwidth]{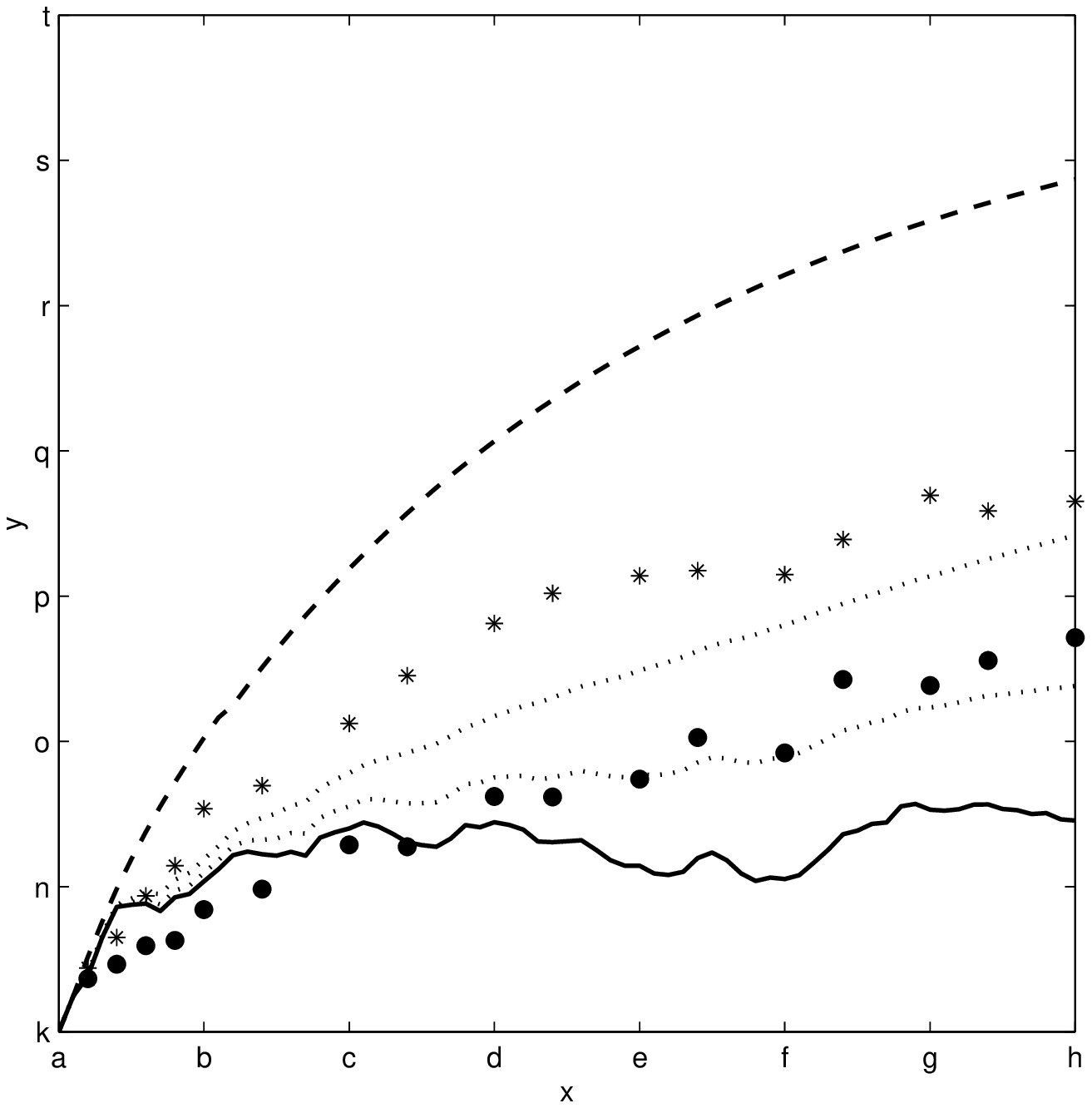}
\vspace{30pt}
\caption{Percentage of particles to cross the
$\rho=\pm10\pi$ cantori. Experimental results for $\eta = 0.0187$
($\bullet$) and $\eta = 0.0503$ (\*) all at a kicking strength of $\ks =
270$. Quantum simulations for the experimental parameters are
shown as dotted lines.  The classical simulation (dashed) and
quantum simulation with zero spontaneous emission (solid) are
shown for comparison}
\label{decohresults}
\end{figure}

Our experimental results support this prediction (see
Figure~\ref{decohresults}).
For a given kick strength, we tuned closer to resonance in steps
from $\delta = 2.8\unit{GHz}$. A lower limit on the detuning is
imposed by the approximation made in our numerical calculations
that the excited state of the atoms can be adiabatically
eliminated (see Section \ref{ghb1}). The upper limit
($2.8\unit{GHz}$) is mainly due to optical power restrictions from
our diode laser - as the detuning increases, the intensity must
also increase to maintain a constant kicking strength. As the
detuning decreased, the percentage of particles outside the
$1^{\unit{st}}$ KAM boundary increased and the experimental curve
rose towards the classical prediction. Our results also show
reasonable agreement with our quantum mechanical simulations which
included the effects of spontaneous emission. The main sources of
error are the measurement of optical beam power and the finite
resolution of the CCD camera.
\section{Conclusion}
Using laser-cooled caesium atoms, we have observed the controlled
decoherence of a real quantum system via coupling to the
environment. This adds to the previous work on decoherence through
the atom optics realization of the $\delta$-kicked rotor and also
the experiments of Haroche \etal \cite{Haroche} and Wineland \etal
\cite{Wineland}. We have demonstrated that the quantum diffusion
rate tends towards the classical rate with an increasing degree of
decoherence. The introduction of decoherence via spontaneous
emission increases the rate of transport of atoms across the
cantori and alters the characteristic shape of the KAM localized
distribution such that it tends towards the classical (uniform)
distribution.

The link between the quantum domain and the familiar classical
world remains a hotly debated topic. The quantum classical
correspondence principle requires that quantum mechanics contains
the classical macroscopic limit. Taking $\hbar \rightarrow 0$ is
not a realistic strategy in the laboratory. The physical effects
of a very small, but definitely non-zero, $\hbar$ manifest
themselves in chaotic systems through highly complex phase space
structures of order $\sim \hbar$ that develop in logarithmically
short times $\sim \ln(1/\hbar)$. It appears that only through
environment induced decoherence can the quantum --- classical
correspondence principle be justified. This interpretation is
authenticated via the results presented in this present paper, as
well as our group's previous study of decoherence in the atomic
optics manifestation of the delta kicked rotor
\cite{Ammann1,Ammann2}. While we can make our ensemble of Cs atoms
behave like an ensemble of classical particles, we do not infer
that there is any degree of chaos in the presented quantum system.
Everything happening to our atoms is intrinsically quantum
mechanical, but through the slight phase shift that an atom's
wavefunction acquires when it recoils from the spontaneous
emission absorption --- emission cycle the momentum localization
is destroyed. This process can be observed and understood in the
laboratory, or through Monte Carlo wavefunction \cite{Ammann1} or
density of state calculations. We can track our theoretical atom's
momentum wavefunction as it absorbs and re-emits environmental
photons, and through this process localization is eliminated and
diffusion similar to that predicted classically is observed. It is
still all quantum mechanics; the physical coupling to the
environment produced behaviour that only ``appears" classical.
\section*{Acknowledgements} This work was supported by the Royal Society of New Zealand
Marsden Fund and the University of Auckland Research Committee.
The authors would like to thank Dan Walls and other members of the
University of Auckland Quantum Optics group for useful
discussions, guidance and encouragement throughout the course of
this research.
\pagebreak
\bibliographystyle{prsty}
\bibliography{kbib,sources}
\end{document}